\documentstyle[12pt]{article}
\newcommand \be {\begin{equation}} \newcommand \beq {\begin{equation}}
\newcommand \bea {\begin{eqnarray} \nonumber } \newcommand \ee
{\end{equation}} \newcommand \eeq{\end{equation}} \newcommand \eea
{\end{eqnarray}} \newcommand{\beqa}{\begin{eqnarray}}
\newcommand{\eeqa}{\end{eqnarray}}

\newcommand{\bc}{\begin{center}} \newcommand{\ec}{\end{center}}

 \def\(({\left(} \def\)){\right)} \def\[[{\left[}
 \def\]]{\right]}

\begin{document}                % End of preamble and beginning of text

\begin{titlepage}
\vskip .27in
\begin{center}
{\large \bf Fluctuation Theorem for stochastic dynamics }

\vskip .27in

{Jorge Kurchan}
\vskip .2in {\it Laboratoire de Physique Th\'eorique de l' Ecole
Normale Sup\'erieure de Lyon \\ All\'ee d'Italie, Lyon, France \\
Jorge.Kurchan@enslapp.ens-lyon.fr} \\ {and} \\ {\it Institute for
Theoretical Physics \\ University of California at Santa Barbara \\
California 93106}
\vskip .2in

\end{center}
\date\today

\vskip 8pt

%\narrowtext

\vspace{0.5cm}
\begin{center}
{\bf Abstract}
\end{center}
\vspace{0.5cm}

The fluctuation theorem of Gallavotti and Cohen holds for finite
 systems undergoing Langevin dynamics. In such a context all
 non-trivial ergodic theory issues are by-passed, and the theorem
 takes a particularly simple form.

As a particular case, we obtain a nonlinear fluctuation-dissipation
theorem valid for equilibrium systems perturbed by arbitrarily strong
fields.

\vspace{1cm}

%{ {\bf \footnotesize
%LPTENS-97/31}}

%\pacs{PACS Numbers~: 75.10.Nr, 64.60.Cn, 64.70.Pf, 11.17.+y}

\end{titlepage}

\newpage

\vspace{.5cm}
\section{Introduction}
\setcounter{equation}{0}
\renewcommand{\theequation}{\thesection.\arabic{equation}}
\label{Introduction}
\vspace{.5cm}

The fluctuation theorem (FT) concerns the distribution of entropy
production over long time intervals. It states that the ratio of the
probabilites of having a given entropy production $\sigma_t$ averaged
over a (large) time interval to that of having $-\sigma_t$ is $e^{t
\sigma_t}$.  It was stated and proved in Refs. \cite{GC} (in what
follows GC) for thermostated Hamiltonian systems driven by external
forces, under certain `chaoticity' assumptions for the dynamics \cite{AA}.

The relevance of this at first sight bizarre result became clear when
 it was shown \cite{G-FDT} that it reduces to the
 fluctuation-dissipation and the Onsager relations in the limit of
 zero power input ({\em i.e.} in equilibrium).

 In this paper we show how to derive the GC fluctuation theorem for
systems undergoing Langevin dynamics.  The purpose of the exercise is
threefold:

\begin{itemize}

\item

The Langevin dynamics is trivially `ergodic', in the sense that for
purely conservative forces, bounded systems with finitely many degrees
of freedom reach the Gibbs-Boltzmann distribution irrespective of the
form of the interaction.

For this reason, one can make a proof of the fluctuation theorem that
is as simple as it can possibly be, having by-passed every non-trivial
question of ergodic theory.  For example, the stationary states are in
this context the zero eigenvalues of a certain (non-Hermitian)
Shroedinger-like operator, and are of a rather familiar nature.

Because of this extreme simplicity, one can use the Langevin systems
 as an heuristic tool to find new results, and after try to see
 whether they hold for more general thermostated Hamiltonian systems.

\item

In order to prove the FT in GC , in addition to making some
 assumptions regarding the `chaoticity' of the models, some conditions
 of boundedness and finitness of the number of degrees of freedom were
 also required.  Here, because all `ergodicity' aspects have been put
 by hand, one can study how the FT can be violated {\em in problems
 with (and as a consequence of) having infinitely many degrees of
 freedom}.  Hence we have a formalism that allows to isolate the
 violations due to `complexity' ({\em i.e.} nontrivial features
 specific to the large-size limit) from violations due to the possible
 non-aplicability of the chaotic hypothesis.

  There are situations for which in the limit of zero forcing the
fluctuation-dissipation theorem is violated in a stationary state {\em
of an infinite system}\cite{driven}.  Because the FT reduces to the
fluctuation-dissipation theorem in the limit of zero non-conservative
forces \cite{G-FDT}, this is a particular instance in which the FT is
violated.

\item

There are many interesting driven systems that can be well represented
by Langevin problems, {\em e.g.} Burgers-KPZ, phase-separation under
shear, turbulence, etc.

\end{itemize}

This paper is organized as follows: in section II we review the
equations describing the evolution of the probability distribution for
a Langevin process. We do this for the case with inertial (second
time-derivative) term, corresponding to the Kramers equations.  We
discuss the (crucial) property of detailed balance, and how it is
modified in the presence of nonconservative forces.

In section III we show that the modified detailed balance propery
leads to the FT. We also present the {\em limit theorem} for the
entropy production \cite{Sinai}, as applied to the stochastic case.
We then follow the steps of Gallavotti \cite{G-FDT} in showing how the
FT reduces to the Green-Kubo formula in the purely conservative limit.

We construct a particular form of the FT corresponding to purely
 conservative driving forces which yields a nonlinear generalization
 of the usual fluctuation-dissipation theorem.

In section IV we show that the arguments of the preceeding sections
 can be applied to a Langevin process without inertial term,
 corresponding to the Fokker-Planck equation.  We present also a
 direct proof of the non-linear fluctuation-dissipation theorem for
 this case.

In the Conclusions we discuss the possible violations of the FT
 equality in systems with infinitely many degrees of freedom.

\pagebreak

\section{Langevin and Kramers Equations}
 
We will consider the Langevin dynamics
\begin{equation}
m {\ddot x}_i + \gamma {\dot x}_i + \partial_{x_i} U({\boldmath x})
+f_i =\Gamma_i,
\label{eq:ksde}
\end{equation} 
where $i=1,\dots,N$.  $\Gamma_i$ is a delta-correlated white noise
with variance $2\gamma T$.

The $f_i$ are velocity-independent forces that do not necessarily
derive from a potential.

We will not deal here with the limits $\gamma=0$ (Hamiltonian
 dynamics), $T=0$ (noiseless dynamics) and $N \rightarrow \infty$, for
 reasons that will become clear.

 We shall first treat in detail the case with inertia $m \neq 0$,
leading to Kramer's equation, and later indicate how to treat the case
in which $m=0$ which leads to Fokker-Planck equation.

\vspace{.5cm}

{\em Kramers equation.}

\vspace{.5cm}

If $m \neq 0$ the probability distribution at time $t$ for the process
(\ref{eq:ksde}) is expressed in terms of the phase-space variables
$x_i$, $v_i$ and is given by \beq P({\boldmath x},{\boldmath v},t)=
e^{ -t H} P({\boldmath x},{\boldmath v},0) \eeq where $H$ is the
Kramers operator \cite{Risken}:
\begin{equation}
H= \partial_{x_i}v_i - \frac{1}{m} \partial_{v_i} \left (\gamma v_i
+ (\partial_{x_i} U({\boldmath x}))+f_i + \gamma \frac{T}{m}
\partial_{v_i} \right)
\label{Kramers},
\end{equation}

We shall find it convenient to express things in bracket and operator
notation:
\begin{eqnarray}
P({\boldmath x},{\boldmath v},t)= \langle x,v|\phi(t) \rangle
\nonumber \\ |\phi(t) \rangle = e^{ - t H} |\phi(0) \rangle
\label{evol} 
\end{eqnarray}
Expectation values of a variable $O(x,v)$ are obtained as:
\begin{equation}
 \langle O(t) \rangle = \langle -| O |\phi(t) \rangle
\end{equation}
where we have defined the flat distribution:
\begin{equation}
 \langle -|x,v \rangle =1 \;\;\;\; \forall \;\;\;\; x,v
\end{equation}
Introducing the Hermitian operators $ {\hat p}_{x_i}$, $ {\hat
p}_{v_i}$ as:
\begin{eqnarray}
  \langle x,v| {\hat p}_{x_i} |\phi(t) \rangle = -i \frac{\partial
    }{\partial x_i} \langle x,v |\phi(t) \rangle \nonumber \\ \langle
    x,v| {\hat p}_{v_i} |\phi(t) \rangle = -i \frac{\partial
    }{\partial v_i} \langle x,v |\phi(t) \rangle
\label{uno}
\end{eqnarray}
the Kramers Hamiltonian reads:
\begin{equation}
H= i v_i {\hat p}_{x_i} - \frac{i}{m} {\hat p}_{v_i} \left( \gamma
v_i + \partial_{x_i} U({\boldmath x}) +f_i + \gamma \frac{iT}{m} {\hat
p}_{v_i} \right)
\label{Kramers1}
\end{equation}
One can explicitate the `conservative' and the `forced' parts of $H$
(the latter being non-conservative if the $f_i$ do not derive from a
potential):
\begin{equation}
H =H^c - \frac{i}{m} {\hat p}_{v_i} f_i
\label{sepak}
\end{equation}

Probability conservation is guaranteed by
\begin{equation}
 \langle -|H=0
\end{equation}
A stationary state satisfies:
\begin{equation}
H |stat \rangle =0 \;\;\;\; ; \;\;\;\; \langle -| stat \rangle =1
\end{equation}

\vspace{.5cm}

{\em Detailed Balance and Time-Reversibility}

\vspace{.5cm}

The evolution of the system satisfies in the absence of
non-conservative forces a form of detailed balance:
\begin{equation}
 \langle x', v'| e^{-t H^c}| x,v \rangle e^{-\beta E_K ({\boldmath
 x},{\boldmath v})} = \langle x, -v| e^{-t H^c}| x',- v' \rangle
 e^{-\beta E_K ({\boldmath x'}, -{\boldmath v}')}
\end{equation}
where the total energy is $E_K({\boldmath x}, {\boldmath v})=
\frac{1}{2} \sum_i m v_i^2 + U({\boldmath x})$.  This leads to a
symmetry property, which in operator notation reads:
\begin{equation}
Q^{-1}_K H^c Q_K = H^{c \; {\dag}}
\label{dbk}
\end{equation}
where the operator $Q_K$ is defined by:
\begin{equation}
Q_K |x,v \rangle \equiv e^{-\beta E_K ({\boldmath x},{\boldmath v})}
|x,-v \rangle
\end{equation}

In the presence of arbitrary forces $f_i$, equation (\ref{dbk}) is
modified to:
\begin{equation}
Q^{-1}_K H Q_K = H^{\dag} + \beta f_i v_i \equiv H^{\dag} -
S^{\dag}
\label{viernes}
\end{equation}
The operator $S$ is the power done on the system divided by the
temperature \beq S^{\dag} = - \beta {\boldmath f} \cdot {\boldmath
v} \eeq and this is the entropy production in the case of a stationary
system. We also have:
\begin{equation}
Q^{-1}_K S Q_K = - S^{\dag}
\label{sabado}
\end{equation}

 Clearly, in the particular case in which the forces $f_i$ derive from
 a potential $f_i= \frac{\partial A}{\partial x_i}$: \beq S= - \beta
 f_i v_i = -\beta \frac{dA}{dt} \eeq whose average value at
 stationarity is zero.

A non-increasing ${\cal H}$-function may be defined as \cite{Kubo}
\begin{equation}
{\cal H}(t)= \int d{\boldmath x} d{\boldmath v} P({\boldmath
x},{\boldmath v}) \left( T \ln P({\boldmath x},{\boldmath v}) +
E({\boldmath x},{\boldmath v}) \right)
\label{H}
\end{equation}
and may be interpreted as a `generalized free-energy'.  Writing
$P_{stat} ({\boldmath x},{\boldmath v}) \equiv \langle x,v|stat
\rangle $, we have that $\dot{{\cal H}} = 0 $ implies that
\begin{equation}
 - \langle - | {\boldmath f} \cdot {\mathbf v} |stat \rangle =
 \gamma \sum_i \int d{\mathbf x} d{\mathbf v} \frac{(mv_i P_{stat}
 +T \partial_{v_i}P_{stat} )^2 }{m^2 P_{stat} } \equiv T \langle
 \sigma \rangle \geq 0
\label{entr}
\end{equation}
 The averaged entropy production at stationarity $ \langle \sigma
 \rangle $ is non-negative
\footnote{ In the pure Hamiltonian $\gamma=0$ case $ \langle \sigma
\rangle =0$ at all times, a consequence of Liouville's theorem.}.

Let us now write a (Green-Kubo) fluctuation-dissipation theorem for a
 {\em purely conservative} system perturbed by an arbitrary small
 force field $- h(t) {\mathbf { f (x)}} $. The current operator is
 \beq J \equiv {\mathbf f} \cdot {\mathbf v} \eeq and $S= - h
 \beta J$.  Linear response theory implies, for any observable
 $O(x,v)$ (cfr. (\ref{evol}) and (\ref{Kramers1})): \beq \frac{\delta
 \langle O(t) \rangle }{\delta h(t')}|_{h=0} =- \frac{i}{m} \langle -|
 O e^{-(t-t')H^c} {\hat p}_{v_i} f_i e^{-t'H^c} |init \rangle
\label{ger}
\eeq In equilibrium $|init \rangle =|GB \rangle $, the Gibbs-Boltzmann
distribution and: \beq -\frac{i}{m} {\hat p}_{v_i} |GB \rangle = \beta
v_i |GB \rangle \;\;\; ; \;\;\; e^{-t'H} |GB \rangle = |GB \rangle
\eeq which introduced in (\ref{ger}) implies the
fluctuation-dissipation theorem: \beq \frac{\delta \langle O(t)
\rangle }{\delta h(t')} \left|_{h=0} = \beta \langle -| O
e^{-(t-t')H^c} \; J | GB \rangle = \beta \langle O(t) \; J(t')
\rangle \right. \theta(t-t')
\label{fdt}
\eeq

%\section{ Modified Detailed Balance and the Fluctuation Theorem}

\vspace{.5cm}

{\em Power and entropy production distribution}

\vspace{.5cm}

Consider the power $ T \sigma_t$ done by the forces $f_i$ in a time
$t$ for a given path in phase space $({\mathbf x} (t'),{\mathbf v}
(t'))$: \beq T \sigma_t \equiv \int_0^t {\mathbf f(x) } (t') \cdot
{\mathbf v} (t') \eeq We wish to study the distribution
$\Pi_t({\sigma_t})$ of $\sigma_t$ for different noise realizations.
Let us show that:
\begin{equation}
\Pi_t({\sigma_t}) = t \int_{-i \infty}^{+i \infty} \; d \lambda \;
 \langle -| e^{-t(H+ \lambda S)} |init \rangle \; e^{t\lambda \sigma_t
 }
\label{cosa}
\end{equation}
 This is most easily seen in the path-integral representation.
 Denoting ${\cal{S}}$ the action associated with $H$ along a path we
 have \beqa \Pi_t({\sigma_t}) & & t \int_{-i \infty}^{+i \infty} d
 \lambda \langle -| e^{-t(H+ \lambda \beta {\mathbf f} \cdot
 {\mathbf v}- \lambda \sigma_t )} |init \rangle \nonumber \\ & &
 \;\;\;\;\; = t \int_{-i \infty}^{+i \infty} \; d \lambda \; \int
 {\cal{D}} (paths) e^{ - {\cal{S}} (path) - \lambda \left(\beta
 \int_0^t {\mathbf f} \cdot {\mathbf v} (t') \; dt' - t \langle
 \sigma \rangle p \right) } \nonumber \\ & & \;\;\;\;\; =t \int
 {\cal{D}} (paths) e^{- {\cal{S}}(path)} \;\; \delta \left(
 \beta\int_0^t {\mathbf f} \cdot {\mathbf v} (t') - t \sigma_t
 \right) \eeqa Bearing in mind that the factor $e^{- {\cal{S}}(path)}$
 is precisely the probability of each path, the last equality implies
 (\ref{cosa}).

 In the next subsection we shall assume that there is a non-zero
average $\langle \sigma \rangle$, and following GC we shall work with
the adimensional variable \beq p=\frac{\sigma_t}{ \langle \sigma
\rangle } \eeq and study the distribution of $p$ given by $\pi_t(p)
\equiv \langle \sigma_t \rangle \Pi_t(\sigma_t p)$.

\vspace{.5cm}

\section{ Modified Detailed Balance and the Fluctuation Theorem}

\vspace{.5cm}

{\em A first version of the Fluctuation Theorem}

\vspace{.5cm}

The fluctuation theorem follows from the modified form of detailed
balance Eqs. (\ref{viernes}) and (\ref{sabado}). These two imply, for
any $\lambda$: \beq Q^{-1} (H + \lambda S) Q= H^{\dag} - (1+ \lambda)
S^{\dag}=[H - (1+ \lambda^*) S]^{\dag}
\label{sime}
\eeq so that $(H + \lambda S)$ and $(H - (1+ \lambda^*) S)$ have
conjugate spectra.  This relation has consequences for the
distribution of power.  Let us see what are the implications for
$\pi_t(p)$, starting from an initial distribution $|init \rangle $.
\beqa \pi_t(p) &=& t <\sigma> \int_{-i \infty}^{+i \infty} \; d
\lambda \; \langle - | e^{-t(H+ \lambda S)} |init \rangle \;
e^{t\lambda \langle \sigma \rangle p} \nonumber \\ &=&t <\sigma>
\int_{-i \infty}^{+i \infty} \; d \lambda \; \langle - | Q e^{-t[H -
(1+ \lambda^*) S]^{\dag}]} Q^{-1} |init \rangle \; e^{t\lambda \langle
\sigma \rangle p} \nonumber \\ &=& t <\sigma> \int_{-i \infty}^{+i
\infty} \; d \lambda \; \langle init| Q^{-1 \; \dag} e^{-t[H - (1+
\lambda^*) S]} Q^{\dag} |- \rangle ^* \; e^{t\lambda \langle \sigma
\rangle p} \eeqa Using that $H$, $S$ and $Q$ are real: \beq \pi_t(p)
=t <\sigma> \int_{-i \infty}^{+i \infty} \; d \lambda \; \langle init|
Q^{-1 \; \dag} e^{-t[H - (1+ \lambda) S]} Q^{\dag} |- \rangle \;
e^{t\lambda \langle \sigma \rangle p} \eeq Making $\lambda \rightarrow
-1-\lambda$: \beq \pi_t(p)=t <\sigma> e^{-t \langle \sigma \rangle p}
\int_{ -1 - i \infty}^{-1+ i \infty} \; d \lambda \; \langle init|
Q^{-1 \; \dag} e^{-t[H + \lambda S]} Q^{\dag} |- \rangle \; e^{t
\lambda \langle \sigma \rangle (-p)}
\label{sep}
\eeq Now, $ \langle A| e^{-t(H+ \lambda S)} |B \rangle $ is for given
$|A \rangle $, $|B \rangle $ an analytical function of $\lambda$, and
we can deform the contour in the integral of the last line to $
\int_{-i \infty}^{+i \infty}$.

Consider firstly the case in which we start from a Gibbs-Boltzmann
distribution $|GB \rangle $ \cite{Pierre}, which {\em need not be} a
stationary distribution in the presence of the nonconservative forces
\footnote{This situation can be experimentally done by switching on
the non-conservative forces at $t=0$ on an equilibrated conservative
system.}.  We then have: \beq \langle GB| Q^{-1 \; \dag} \propto
\langle -| \;\;\;\; ; \;\;\;\; Q^{\dag} |- \rangle \propto |GB \rangle
\eeq and Eq. (\ref{sep}) implies, {\em for all times}: \beqa
\pi^{GB}_t(p) &=& e^{-t \langle \sigma \rangle p} \;\; \pi^{GB}_t(-p)
\\ - \langle \sigma \rangle p &=& \frac{ \ln(\pi^{GB}_t(p)) -
\ln(\pi^{GB}_t(-p))}{t}
\label{oct}
\eeqa Here we have added the superscript $GB$ to indicate the initial
condition.

What about other initial conditions? Already at this stage it becomes
intuitive that if the system is such that any initial condition (in
particular the Gibbs-Boltzmann distribution) evolves {\em in finite
time} $\sim \tau_{erg}$ to the same stationary distribution $|stat
\rangle $, then we should have a result like (\ref{oct}) (but only for
$t >> \tau_{erg}$) irrespective of the initial condition.  Note that,
surprisingly, the stationary distribution does not appear to play a
special role here, but the Gibbs-Boltzmann distribution does!
However, this statement has to be qualified if we wish to identify the
power done by the nonconservative forces as an `entropy production',
something we can do only in the stationary regime.

\pagebreak

\vspace{.5cm}

{\em Long-time distributions of} $\pi(p)$.

\vspace{.5cm}

Let us make the remarks above more precise. If, under certain
assumptions we have that for long times there is a single limiting
function $\zeta(p)$ \cite{Sinai} such that {\em irrespective of the
initial conditions}
\begin{equation}
\lim_{t \rightarrow \infty} \frac{1}{ \langle \sigma \rangle t} \; \ln
(\pi_t) \rightarrow - \zeta(p)
\label{cosa1}
\end{equation}
 then the FT (\ref{oct}) will hold for long times for any initial
condition, and will read: \beq \zeta(p)-\zeta(-p)= \langle \sigma
\rangle p
\label{FT}
\eeq

In order to derive (\ref{cosa1}) we shall make the two following
assumptions:

{\em i} The lowest (zero) eigenvalue of $H$ is non-degenerate.

{\em ii} The initial state has a non zero overlap with the left
eigenvector of $H$; \newline $\langle stat|init \rangle \neq 0$.

{\em Any of these assumptions may fail to hold in unbounded systems or
 in systems with infinitely many degrees of freedom}. Indeed,
 conservative systems with slow dynamics such as glasses and
 coarsening are known to have a gap-less spectrum of the Fokker-Planck
 operators (the gap goes to zero with the system's number of degrees
 of freedom).

Furthermore, the gap vanishes in the purely Hamiltonian $\gamma=0$
limit, as there are many long-lived phase-space distribution in that
case ({\em e.g.} invariant tori, etc), as well as in the $T=0$
case. Note that $H$ loses the second derivative in these cases.  This
is the main reason why we only consider $\gamma>0, T>0$ here.

We proceed as follows: Introducing the right and left eigenvectors:
\beq (H+ \lambda S) |\psi_i^R(\lambda) \rangle = \mu_i(\lambda)
|\psi_i^R(\lambda) \rangle \;\;\;\; ; \;\;\;\; \langle
\psi_i^L(\lambda)|(H+ \lambda S) = \mu_i (\lambda) \langle
\psi_i^L(\lambda)|
\label{eigenvectors}
\eeq we have:
\begin{equation}
\pi^{|init \rangle }_t(\sigma_t) = t <\sigma> \int_{-i \infty}^{+i
 \infty} \; d \lambda \; \sum_i \; \langle -|\psi_i^R(\lambda) \rangle
 \langle \psi_i^L (\lambda)|init \rangle e^{-t(\mu_i(\lambda)-\lambda
 \langle \sigma \rangle p)}
\end{equation}

Let us denote $\mu_0(\lambda)$ the eigenvalue with lowest real part
and the corresponding left and right eigenvectors $ |\psi_0^R(\lambda)
\rangle $ and $ |\psi_0^L(\lambda) \rangle $.  Under the assumptions
above, there will be at least a range of values of $\lambda$ around
zero such that the eigenvalue $\mu_0(\lambda)$ will be nondegenerate.
Then, the integral over $\lambda $ will be dominated for large $t$ by
the saddle point value:
\begin{equation}
 \zeta(p) = \mu_0(\lambda_{sp})- \lambda_{sp} \langle \sigma \rangle p
\label{cosa2}
\end{equation}
where the saddle point $\lambda_{sp}$ is a function of $p$ determined
by:
\begin{equation}
 \frac{d \mu_0}{d\lambda} (\lambda_{sp})= \langle \sigma \rangle p
\label{cosa3}
\end{equation}
The dependence upon the initial distribution is, within these
assumptions, subdominant for large $t$.

In order to check that the distribution of $p$ obtained from
(\ref{cosa2}) and (\ref{cosa3}) is indeed peaked at $p=1$, we
calculate
\begin{equation}
\frac{d \zeta (p)}{d p} = \left. \left.\left. \frac{d \mu_0}{d
 \lambda}\right|_{\lambda_{sp}} \frac{d \lambda}{d
 p}\right|_{\lambda_{sp}} - \frac{d\lambda}{d p}\right|_{\lambda_{sp}}
 \langle \sigma \rangle p - \langle \sigma \rangle \lambda_{sp}= -
 \langle \sigma \rangle \lambda_{sp}
\end{equation}
Hence, the minimum of $\zeta$ happens at $\lambda_{sp}=0$.  We can
further calculate the derivative of an eigenvalue using first order
perturbation theory. Eq. (\ref{cosa3}) for $\lambda_{sp}=0$ then
reads: \beq \langle \sigma \rangle p|_{\lambda_{sp}=0}= \frac{d
\mu_0}{d\lambda}|_{\lambda_{sp}=0}= \frac{ \langle \psi_0^L(0)| S
|\psi_0^R(0) \rangle }{ \langle \psi_0^L(0)|\psi_0^R(0) \rangle } =
\langle -| S |stat \rangle = \langle \sigma \rangle \eeq where we have
used that in the absence of perturbation the lowest eigenvalue is zero
and corresponds to the (unique) stationary state. Since then
$\mu_0(0)=0$, we obtain that $\zeta(p)$ takes its minimum (zero) value
at precisely $p=1$ ({\em i.e. } $\sigma_t= \langle \sigma \rangle $),
and for large $t$ the distribution is sharply peaked at $p=1$, as it
should.
 
\vspace{.5cm}

{\em The Fluctuation -Dissipation Theorem as the `conservative limit'
limit of the FT}

\vspace{.5cm}

Gallavotti has shown \cite{G-FDT} that the FT gives the
fluctuation-dissipation theorem in the conservative limit of zero
entropy production.  Here, we shall paraphrase that derivation, as
applied to Langevin case.

Before doing that, let us first notice that the detailed balance
symmetry (\ref{viernes}) is responsible in the purely conservative
$S=0$ case for the existence of fluctuation-dissipation and
reciprocity relations.  As we have seen, it is also responsible in the
driven $S \neq 0$ case for the FT relation.  The result in Ref.
\cite{G-FDT} is that the FT formula is {\em on its own} enough to give
us back the fluctuation-dissipation and the reciprocity relations in
the purely conservative limit.

Let us rewrite a form of the fluctuation-dissipation theorem for a
conservative systems.  In equation (\ref{fdt}) we set $O=J$, and
compute the response to a force $- h {\mathbf f}$ constant in time.
Integrating (\ref{fdt}) over $t,t'$: \beq \int_0^t \;
dt'\frac{\partial \langle J (t')\rangle }{\partial h}|_{h=0} = \beta
\int_0^t \; dt' \; dt'' \; \langle -| J e^{-(t'-t'')H^c} \; J | GB
\rangle \theta(t-t')
\label{fdt3}
\eeq The right hand side of (\ref{fdt3}) can be reexpressed as
follows: \beq \frac{ \beta}{2} \int_0^t \; dt' \; dt'' \;\langle -|
e^{-(t-t')H^c} J e^{-(t'-t'')H^c} \; J e^{-t''H^c}| GB \rangle =
\frac{ 1}{2\beta } \frac{\partial^2 }{\partial \lambda ^2} \langle -|
e^{-t(H^c+\lambda \beta J)}| GB \rangle\left|_{\lambda=0} \right.
\eeq where we used that $H^c$ anihilates both $\langle -|$ and $|GB
\rangle$.  Similarly, the left hand side of (\ref{fdt3}) can be
reexpressed as \beqa \frac{\partial }{\partial h} \int_0^t \; dt'
\langle -| J e^{-t'H(h)} |GB>\left|_{h=0} \right.  &=& \frac{\partial
}{\partial h} \int_0^t \; dt' \langle -| e^{-(t-t')H(h) } J
e^{-t'H(h)} |GB>\left|_{h=0} \right. \nonumber \\ &=& \frac{1}{\beta}
\frac{\partial^2}{\partial h \partial \lambda} \langle -| e^{-t(H(h)+
\lambda \beta J) } |GB \rangle \left|_{\stackrel{h=0}{\lambda=0}}
\right.  \eeqa where $H(h)$ is the perturbed Hamiltonian.  Then, we
can rewrite the fluctuation-dissipation theroem as: \beq \left[
\frac{1}{2} \frac{\partial^2 }{\partial \lambda^2} -
\frac{\partial^2}{\partial h \partial \lambda} \right] \langle -|
e^{-t(H(h) +\lambda \beta J)}| GB \rangle
\left|_{\stackrel{h=0}{\lambda=0}}\right. = 0
\label{fdtfdt}
\eeq Using the fact that $\langle J \rangle=0$ in equilibrium for a
conservative system, the first derivatives with respect to $\lambda$
and with respect to $h$ vanish and Eq. (\ref{fdtfdt}) can be rewritten
as: \beq \left[ \frac{1}{2} \frac{\partial^2 }{\partial \lambda^2} -
\frac{\partial^2}{\partial h \partial \lambda} \right] \ln \left\{
\langle -| e^{-t(H(h) +\lambda \beta J)}| GB \rangle \right\}
\left|_{\stackrel{h=0}{\lambda=0}}\right. = 0
\label{fdtfdt1}
\eeq To lowest (quadratic) order in $h$ and $\lambda$, the general
solution of (\ref{fdtfdt1}) is: \beq \ln \left\{ \langle -| e^{-t(H(h)
+\lambda \beta J)}| GB \rangle \right\} = A(t) \left( \lambda^2 +
(\lambda + h)^2 \right) + B(t) \lambda (h + \lambda)
\label{cosas}
\eeq where $A,B$ are model-dependent. Equation (\ref{cosas}) is a form
of the fluctuation-dissipation theorem.

Let us now show that the FT directly implies (\ref{cosas}) in the
purely conservative ($h=0$) limit.  We have that: \beq \ln \left\{
\langle -| e^{-t(H(h) +\lambda \beta J)}| GB \rangle \right\}= \ln
\left\{ \langle -| e^{-t(H(h) - \frac{\lambda}{h} S)} |GB \rangle
\right\} = \int_{-i \infty}^{+i \infty} \; dp \; \pi(p) e^{\frac{-t
\langle \sigma \rangle p\lambda}{h}} \eeq Now, {\em with the only
assumption of the FT} applied to the term on the right we easily
obtain: \beq \ln \left\{ \langle -| e^{-t(H(h) + \frac{\lambda}{h} S)}
|GB \rangle \right\} = \ln \left\{ \langle -| e^{-t(H(h) -
(1+\frac{\lambda}{h}) S} |GB \rangle \right\} \eeq which to second
order in $\lambda,h$ implies (\ref{cosas}).

\vspace{.5cm}

{\em A non-linear Fluctuation -Dissipation Theorem in the
`conservative limit' }

\vspace{.5cm}

So far we have only concentrated on the case in which the forces $f_i$
do not derive from a potential and thus generate entropy at
stationarity.  However, the calculations can be performed in the case
in which the $f_i$ do derive from a potential: \beq f_i = -h
\frac{\partial A( {\mathbf x})}{\partial x_i} \eeq In that case, $S$
is no longer an entropy production, but it represents the rate of
variation of $A$.  For example: \beq S= \beta h \frac{\partial A(
{\mathbf x})}{\partial x_i} v_i = \beta h \frac{d A}{dt} \eeq If we
start at $t=0$ with the equilibrium distribution {\em in the absence
of forces $f_i$} and only then switch on the forces, we easily obtain
a version of the FT: \beq
\frac{\pi_h(A(t)-A(0)=a)}{\pi_h(A(t)-A(0)=-a)} = e^{ \beta h a}
\label{obt}
\eeq valid for {\em arbitrary} $h$. Here the subindex $h$ means that
the distribution depends on the field conjugate to $A$ that has been
on from $t=0$ to $t$.

Using (\ref{obt}) in the limit of small $h$, we recover the usual
fluctuation-dissipation theorem as follows: \beqa \langle A(t) - A(0)
\rangle|_h &=& \int a \; da \; \pi_h(A(t)-A(0)=a) = - \int a \; da \;
\pi_h(A(t)-A(0)=-a) \nonumber \\ &=& - \int da \; e^{-\beta h a} a \;
\pi_h(A(t)-A(0)=a) \eeqa This yields, to lower order in $h$: \beqa
\langle A(t) - A(0) \rangle|_h &=& - \int a \; da \;
\pi_h(A(t)-A(0)=a) (1-\beta h a) \nonumber\\ &=&- \int a \; da \;
\pi_h (A(t)-A(0)=a) \nonumber \\ & & - h \beta \int a^2 \; da \;
\pi_{h=0}(A(t)-A(0)=a) \eeqa which implies: \beq \frac{d}{dh} \langle
A(t) - A(0) \rangle|_{h=0} = \frac{\beta}{2} \langle (A(t) - A(0))^2
\rangle|_{h=0} \eeq which is a usual form of the
fluctuation-dissipation theorem.

\pagebreak

\vspace{.5cm}

\section{Fokker-Planck equation}

\vspace{.5cm}

If the inertial term in the Langevin equation vanishes the probability
distribution at time $t$ for the process (\ref{eq:ksde}) can be
expressed only in terms of the positions, (in fact, the velocities are
in this case undefined). The evolution is now given by \beq
P({\mathbf x},t)= e^{ -t H_{FP}} P({\mathbf x},0) \eeq where
$H_{FP}$ is the Fokker-Planck operator:
\begin{equation}
H_{FP}= -\partial_{x_i} \left( T \partial_{x_i} + \partial_{x_i}
U({\mathbf x}) + f_i \right)
\label{Fokker}
\end{equation}
We have set $\gamma=1$ for this case.  In bracket notation, we have:
\begin{eqnarray}
P({\mathbf x},t) &=& \langle x|\phi(t) \rangle \nonumber \\ |\phi(t)
\rangle &=& e^{ - t H_{FP}} |\phi(0) \rangle
\end{eqnarray}

\begin{equation}
H_{FP}= {\hat p}_{x_i} \left( T {\hat p}_{x_i} - i \partial_{x_i} U -
i f_i\right)
\label{Fokker1}
\end{equation}
which can again be separated in conservative and `forced' parts:
\begin{equation}
H_{FP}= H_{FP}^c -i {\hat p}_{x_i} f_i
\end{equation}
The total energy of the conservative part for this case is simply $U$.

The evolution of the system satisfies the usual detailed balance {\em
in the absence of non-conservative forces}:
\begin{equation} 
 \langle x'| e^{-t H_{FP}^c}| x \rangle e^{-\beta U ({\mathbf x})} =
 \langle x| e^{-t H_{FP}^c}| x' \rangle e^{-\beta U ({\mathbf x'})}
\end{equation}
leading to:
\begin{equation}
Q^{-1}_{FP} H_{FP}^c Q_{FP}= H^{c \; \dag}_{FP}
\label{dbf}
\end{equation}
where the operator $Q_{FP}$ is defined by:
\begin{equation}
Q_{FP} |x \rangle \equiv e^{-\beta U ({\mathbf x})} |x \rangle
\end{equation}

\vspace{.5cm}

{\em Power and entropy production for Fokker-Planck processes}

\vspace{.5in}

In a Langevin process without inertia it is not {\em a priori} obvious
how to define the power done by the bath. In order to do this, we
shall first make a heuristic argument and only then formalise it.  Let
us compute the power as 
\beq T \sigma_t = \int_0^t dt \; {\dot x}_i
f_i 
\eeq
 Because the velocity is not well defined, we shall have to be
careful about the meaning of this expression. Let us for the moment
continue naively, writing a functional expression for the distribution
$\Pi(\sigma_t)$:
 \beqa 
\Pi(\sigma_t) &=& t \left[ \int D({\mathbf x}) \;
{\mathbf J(x)} \; \delta({\dot x}_i + \partial_{x_i} U({\mathbf x}) +f_i
-\Gamma_i) \; \delta(t\sigma_t- \beta \int_0^t dt \; {\dot x}_i f_i)
\right]_\Gamma \nonumber \\ = &t& \left[ \int_{-i \infty}^{+i \infty}
d \lambda \int D({\mathbf x}) \; D({\mathbf p}) {\mathbf J(x)} e^{-\int_0^t [ i
p_i({\dot x}_i + \partial_{x_i} U({\mathbf x}) +f_i -\Gamma_i)
+\lambda \beta \; {\dot x}_i f_i}]\right]_\Gamma \; e^{\lambda t
\sigma_t} \nonumber
 \eeqa
 where ${\mathbf J}$ is the Jacobian associated
with the equation of motion delta.  The square brackets denote
averaging over the noise.  Performing this average, we obtain:
 \beqa
\Pi(\sigma_t) &=& t \int_{-i \infty}^{+i \infty} d \lambda \int D({\mathbf
x}) \; D({\mathbf p}) \; {\mathbf J(x)} \; e^{-\int_0^t dt' i p_i ( {\dot x}_i
+ \partial_{x_i} U({\mathbf x}) +f_i -i T p_i)} \; \nonumber \\ & &
e^{-\lambda \int_0^t dt' (2 i f_i p_i - \beta ( f_i + \partial_{x_i}
U({\mathbf x})) - \lambda^2 \beta \int_0^t dt'  f_i^2)} \;
e^{\lambda t \sigma_t} \nonumber
\label{vu}
\eeqa

Equation (\ref{vu}) is a functional expression of the equality: \beq
\Pi(\sigma_t) =t \int_{-i \infty}^{+i \infty} d \lambda \langle -|
e^{-t[H_{FP} - \lambda ( i f_i ( {\hat p}_{x_i} - i \beta \partial_{x_i}
U({\mathbf x}) - i \beta f_i) + i {\hat p}_{x_i} f_i ) - \beta \lambda^2 f_i^2 ]}
|init \rangle \; e^{t\lambda \sigma_t }
\label{cosa'}
\end{equation}
This expression, to be compared with (\ref{cosa}), can be taken as the
definition of `power done by the bath', yielding the entropy
production at stationarity.  Note that we have made in this definition
a precise choice of factor orderings (we have associated 
 the c-number $2p_i f_i$ with the operator
 ${\hat p}_{x_i} f_i +  f_i {\hat p}_{x_i}$).

We are now in a position of exploring the consequences of a modified
 detailed balance.  Proposing a tranformation like (\ref{dbf}) we find
 that: 
\beqa 
& & e^{\beta U} [H_{FP} - \lambda ( i f_i ( p_i - i \beta
 \partial_{x_i} U({\mathbf x}) - i \beta f_i) + i p_i f_i ) - \beta
 \lambda^2 f_i^2 ] e^{-\beta U} = \nonumber \\
 & & [H_{FP} - {\tilde
 \lambda} ( i f_i ( p_i - i \beta \partial_{x_i} U({\mathbf x}) - i
 \beta f_i) + i p_i f_i ) - \beta {\tilde \lambda}^2 f_i^2 ]^{\dag}
 \eeqa
 where ${\tilde \lambda} \equiv -(1+\lambda)$.  Performing the
 change of variables $ \lambda \to {\tilde \lambda}$ in the integral
 and following the same steps as in the Kramers case we obtain the
 fluctuation theorem for $\Pi(\sigma_t)$ defined as in (\ref{cosa'}).

\vspace{.5cm}

{\em Nonlinear fluctuation-dissipation theorem}

\vspace{.5cm}

 A direct derivation of the nonlinear fluctuation-dissipation theorem
(\ref{obt}) can be made for dynamics which (like Fokker-Planck's, heat
bath, etc) satisfies detailed balance.  Denoting $H(h)$ the evolution
Hamiltonian associated with a potential $U - h A(x)$ we then have:
\beq e^{\beta(U-hA)} H(h) e^{-\beta(U-hA)} =H^{\dag} (h) \eeq
 We write
\beq \pi_h(A(t)-A(0)=a) = \int d \lambda \langle -| e^{\lambda A}
e^{-tH(h)} e^{-\lambda A} | GB \rangle e^{-\lambda a }
\label{bbbbb}
\eeq where $| GB \rangle$ is the canonical distribution {\em at }
$h=0$.  Now, \beqa \langle -| e^{\lambda A} e^{-tH(h)} e^{-\lambda A}
| GB \rangle &=& \langle -| e^{\lambda A} e^{-\beta (U - h A)}
e^{-tH^{\dag}(h)} e^{\beta (U - h A)} e^{-\lambda A} | GB \rangle
\nonumber \\ &=& \langle -| e^{(\beta h +\lambda) A} e^{-tH(h)}
e^{-(\beta h +\lambda) A} | GB \rangle \eeqa Introducing this in
(\ref{bbbbb}) and performing the integral over $\lambda$ one readily
obtains (\ref{obt}).

\section{Conclusions}

In this paper we have derived the FT for Langevin processes of systems
with finitely many degrees of freedom.  We do not require any
properties of the potential apart from boundedness, since the Langevin
equation is `as ergodic as possible'.

However, we have noted in several places that the derivations do not
carry through automatically if the zero eigenvalue of the operator $H$
is degenerate.  This happens surely in the case that the system is
disconnected from the bath ($\gamma=0$) and in the $T=0$ case.  More
interestingly, it may also happen in an infinite $N=\infty$ system.

Indeed, because the `gap' in the spectrum is the inverse of a
time-scale, a gapless spectrum is an indication of `slow' dynamics.
This suggests that the FT might be violated for those (infinite)
driven systems which in the absence of drive have a slow relaxational
dynamics that does not lead them to equilibrium in finite times (as is
the case of glassy systems, coarsening, etc.).

In such systems it is known\cite{driven} that the
fluctuation-dissipation theorem can be violated even in the limit of
vanishing power input, although the frequency range of the violation
respects some bounds \cite{Fdt}.  The violation of the
fluctuation-dissipation theorem (or alternatively, the appearence of
`effective temperatures' different from the bath temperature) seems to
be a signature of the dynamics of conservative or near-conservative
complex systems \cite{Cukupe,review}.  This raises the intriguing
possibility that the violation of FT might exist and play a similar
role for strongly driven infinite systems.

\vspace{2cm}

I wish to thank L.F. Cugliandolo and P. LeDoussal for useful
suggestions.  This work was supported in part by Natural Science
Foundation under grant number PHY9407194.

\end{document}